\documentclass[preprint,12pt]{article}
\usepackage{authblk}

\usepackage{braket}
\usepackage[square, sort&compress, numbers]{natbib}
\usepackage{color}
\usepackage{amssymb,amsmath,amsmath,amsfonts,amsthm,bm}
\usepackage{hyperref}
\usepackage{xcolor}
\hypersetup{
    colorlinks,
    linkcolor={red!75!black},
    citecolor={blue!75!black},
    urlcolor={blue!75!black}}
\usepackage[utf8]{inputenc}
\usepackage{physics}
\usepackage{graphicx}
\usepackage{caption}
\usepackage{subcaption}
\usepackage{tabularx}
\usepackage{enumitem}
\usepackage{cleveref}
\usepackage{threeparttable}
\usepackage{lscape}
\usepackage{natbib}
\usepackage{booktabs}
\graphicspath{{images/}}
\usepackage{cancel}
\usepackage[normalem]{ulem}
\usepackage{soul}

\usepackage{flushend}
\usepackage{color}
\usepackage{float}
\usepackage{multirow}

\definecolor{red}{rgb}{1.0,0.0,0.0}
\newcommand{\ie}{{\it i.e., }}

\usepackage{hyperref}

\begin{document}
\title{\bf{Kaon-meson coupling from SU(3) flavour symmetry and application to antikaon condensed dense matter in neutron star}}   

\author[1]{Athira S.\thanks{\texttt{p23ph0002@iitj.ac.in}}}
\author[1]{Monika Sinha\thanks{corresponding author: \texttt{ms@iitj.ac.in}}}
\author[2]{Vivek Baruah Thapa\thanks{\texttt{vivek.thapa@bacollege.ac.in}}}
\author[3]{Vishal Parmar\thanks{\texttt{vishal.parmar@pi.infn.it}}}

\affil[1]{\small Indian Institute of Technology Jodhpur, Jodhpur 342037 India}
\affil[2]{\small Department of Physics, Bhawanipur Anchalik College, Barpeta, Assam 781352, India}
\affil[3]{INFN, Sezione di Pisa, Largo B. Pontecorvo 3, I-56127 Pisa, Italy}
\maketitle

\begin{abstract}
Observations of massive pulsars suggest that the central density of neutron stars can exceed several times the nuclear saturation density, creating a favourable environment for the appearance of exotic states, such as strange and non-strange baryons, meson condensates, and deconfined quark matter. The antikaon condensate is the most studied and plausible candidate among meson condensates. However, little is known about the exact interaction mechanisms between antikaons and mediator mesons. In this work, we investigate these interactions by employing SU(3) flavour symmetry for the first time to study antikaon condensation in dense matter. We determine hadron couplings in the mesonic sector using SU(3) flavour symmetry. We consider two scenarios involving three key parameters: the mixing angle $\theta_v$ (between octet meson $\omega_8$ and singlet meson $\phi_1$), the octet to singlet coupling ratio $z$, and the symmetric–antisymmetric weight factor $\alpha_v$. We consider two scenarios: in the first, the mesonic sector parameter $\alpha_m$ is varied with fixed $z_m$, comprising the cases where the baryonic sector parameter $\alpha_b = \alpha_m$, $2\alpha_m$, and $\alpha_m/2$; in the second, $z_m$ is varied with fixed $\alpha_m$, comprising the cases where the baryonic sector parameter $z_b = z_m$, $ 2z_m$, and $z_m/2$, to independently explore their respective impacts on the system. Using this approach, we obtain the couplings of antikaons with both octet and singlet vector mesons and examine the resulting implications for dense matter. Our results show that the equation of state becomes increasingly stiffer with higher values of $\alpha_v$ or $z$, delaying the onset of antikaon condensation and raising the maximum mass of neutron stars. These findings demonstrate the sensitivity of the equation of state to meson–antikaon interactions and establish SU(3) symmetry as a robust framework for constraining them.

\end{abstract}

\maketitle

\section{Introduction}
\label{introduction}
Massive stars end their lives by supernova explosion with the core collapsing into either a neutron star (NS) or a black hole, depending on the progenitor star's mass \cite{Glendenning:1997wn, Bandyopadhyay:2021gnp}. If the collapsed core is less than approximately $2.5~M_\odot$, then the core eventually becomes a NS. The NS is very compact, with a mass of around $1-2~M_\odot$ and a radius of around $10-13$ km. Within this compact object, the average density of the matter is $\sim~10^{14}$ gm cm$^{-3}$. The density of the matter inside NS gradually increases from the surface to the centre, spanning sub-nuclear to super-nuclear densities. At the outer crust, matter consists of a mixture of ions and electrons. With the increase of density within the NS crust, the ions consist of neutron-rich nuclei as the outer electrons are pushed into the nuclei and combine with protons to form neutrons. Moving inward in the inner crust, as the density increases and reaches the neutron drip threshold, neutrons begin to escape from nuclei, leading to a composition of free neutrons, neutron-rich nuclei, and electrons. At densities approaching and exceeding nuclear saturation in the deeper regions, the composition shifts to predominantly free neutrons, accompanied by a small proportion of protons and electrons in $\beta$-equilibrium. This is the beginning of the outer core. Recent astrophysical observations of compact stars indicated the existence of highly massive compact stars with masses close to and above $2M_\odot$, such as  
PSR J1614-2230 (M = 1.97 ± 0.04$M_\odot$) \cite{2010ApJ...724L.199O},
PSR J0740+6620 (M = 2.08 ± 0.07$M_\odot$ with 95\% credibility) \cite{2021ApJ...918L..28M},PSR J0348+042 (M = 2.01 ± 0.04$M_\odot$ ) 
\cite{Bandyopadhyay:2021gnp,2018ChJPh..56..292H}, PSR J1810+1744 (M = 2.13 ± 0.04$M_\odot$ ) \cite{2021ApJ...908L..46R}, and PSR J0952-0607 (M = 2.35 ± 0.17$M_\odot$ ) \cite{2022ApJ...934L..17R}. This indicates the density of the inner core may exceed $2$ times the nuclear saturation density $n_0$. At that much density, the composition and properties of matter are still unknown, as matter at such high densities cannot be produced in any terrestrial laboratory. 

From a theoretical point of view, at high densities, there is the possibility of the appearance of exotic degrees of freedom. For example, at such high density, the heavier strange and non-strange baryons \cite{2020Parti...3..660T, 2023PrPNP.13104041S, 2007PrPNP..59...94W, 2011PhRvC..83b5805W, 2018qcs..confa1043M,2025ApJ...980...54L,2024arXiv241201201J,2024PhRvD.110k4009X,2024arXiv240214288W,2021EPJA...57..216D, 2023PhRvC.107d5804M}, the bosons condense like anti-kaons 
\cite{2024PhRvC.110d5804P,2001PhRvC..64e5805B} or the deconfined strange quark matter \cite{1999PhRvC..60a5802P, 1992ApJ...400..647G, 1996PhR...264..143G, 2011PhRvD..83b5012L,2010PhRvD..82f5017L, 2011JKPS...59.2114H} may appear.

The behaviour of highly dense matter containing exotic components remains largely uncertain, as terrestrial experiments cannot directly constrain it. The properties of matter depend on the constituent exotic particles and the interactions among these particles.

One of the possibilities is the appearance of antikaon condensation at high-density regime of the matter \ie in the core of the NS. Kaplan and Nelson first predicted kaon condensation in dense matter within a chiral perturbative framework \cite{1988NuPhA.479..273K, 1987PhLB..192..193N, 1995NuPhA.585..401L}, studies using other models and advancements were also done \cite{1995PhRvC..52.3470K, 1994NuPhA.567..937B, 1994PhLB..326...14L, 2024AcPPB..55....1T,2006PhRvC..73c5802M,2022PhRvC.105a5807M, 1994NuPhA.572..693T}. Veselsky's study discussed the ultralight compact object observed in the supernova remnant HESS J1731- 347 as evidence supporting the hypothesis of an exotic core in neutron stars. This object is analysed using a kaon condensate model in nuclear matter, which softens the hadronic EOS. Veselsky employed two theoretical frameworks to account for various nuclear models and their implications for the kaon condensate \cite{2024arXiv241005083V}. Several researchers have investigated antikaon condensation in nuclear matter \cite{1999PhRvC..60b5803G, 2001PhRvC..64e5805B, 1997PhR...280....1P, 1996PhRvC..53.1416S, 2001PhRvC..63c5802B, 2021EPJST.230..561M, 2020PhRvD.102l3007T, 2010IJMPD..19.1553M}. In previous studies \cite{1999PhRvC..60b5803G,1996PhR...275..255L,1998PhRvL..81.4564G}, the kaon coupling with mediator vector mesons in the presence of antikaon condensation was calculated using the quark model and isospin counting rule (QMIC )\cite{1994AnPhy.235...35S,2001PhRvC..64e5805B,2021ApJ...910...96M,  1963RvMP...35..916D,  2011PhRvC..84c5809R, 2014PhRvC..90a5801C}.
Unlike SU(6) spin-flavour symmetry, which imposes rigid constraints on coupling constants that often lead to the premature appearance of hyperons and an excessively soft equation of state, SU(3) allows for controlled symmetry breaking and more accurate adjustment of hyperon couplings to match known hypernuclear potentials. In the work done by Lopes \textit{et al.} \cite{2014PhRvC..89b5805L}, SU(3) symmetry is shown to enable the determination of hyperon potentials directly from group-theoretical properties, without relying on additional phenomenological inputs, in contrast to the Glendenning conjecture (GC), which uses empirical assumptions to fix couplings. Furthermore, SU(3) symmetry is unified with SU(6) to calculate meson couplings for both spin-1/2 baryon octet and spin-3/2 decuplet states by Lopes \cite{2023PhRvD.107c6011L}, but it is the SU(3) symmetry constraints that play a decisive role in governing the particle composition in neutron star matter. This approach significantly reduces arbitrariness in hyperon-meson couplings and leads to a more theoretically grounded and constrained modelling of dense stellar interiors. Hence, in our approach, we employ SU(3) flavour symmetry in calculating antikaon couplings for neutron star matter. It aligns more closely with QCD principles and provides the necessary flexibility to fit empirical data.

In this work, for the first time, SU(3) symmetry is used for the calculation of hadron couplings in the mesonic sector, and it is extended to study the antikaon condensation in dense nuclear matter. This work estimates antikaon couplings based on SU(3) symmetry in flavour space. The couplings are determined using Clebsch--Gordan coefficients, with only one free parameter. Then we employ the RMF approach with density-dependent couplings in the nucleonic sector to calculate the matter properties with antikaon condensation from SU(3) couplings. With this matter, we construct the NS properties and observe whether they are consistent with the recent astrophysical constraints. In our previous work \cite{2024PhRvC.110d5804P}, we investigated antikaon condensation in neutron star matter using meson couplings derived from SU(6) spin-flavour symmetry. It was found that kaon condensation occurred only in neutron stars exceeding $1.8\, M_\odot$. This result, driven by the relatively soft equation of state (EOS) produced under SU(6) assumptions, highlighted the sensitivity of antikaon onset to the underlying symmetry structure. However, the meson-kaon coupling strengths are not well constrained due to the lack of direct experimental data \cite{2020PhRvD.102l3007T, 2021PhRvD.103f3004T, 2002PhRvC..66f5801B, 2021EPJST.230..561M, 1995NuPhA.585..401L, 2000NuPhA.674..553P}, making it essential to examine alternative frameworks. SU(3) flavour symmetry offers a broader and more flexible structure for these couplings. In this work, we therefore extend our previous study by exploring the implications of SU(3)-based meson couplings for antikaon condensed matter in neutron stars.

The matter structure and interaction of the constituent particles can be modelled in different ways. The most discussed and trivial model is the relativistic mean field (RMF) approach  \cite{1974AnPhy..83..491W}. In this approach, the constituent particles interact via mediating mesons - scalar and vector mesons and the nucleon-meson couplings are determined from the nuclear matter properties at the nuclear saturation density \cite{1994AnPhy.235...35S, 1985ApJ...293..470G, 1995PhRvC..52.3470K}. However, the interaction of other exotic particles, such as strange and heavier non-strange baryons and mesons, remains poorly characterised due to the lack of experimental and theoretical constraints.

The structure of the paper is as follows. Section~\ref{sec:rmf} briefly discusses the RMF formalism. In Section~\ref{sec:SU3}, we present a detailed calculation of the antikaon couplings with mediator mesons within the SU(3) framework and determine the relevant parameters under two different scenarios. Section~\ref{sec:appli} outlines the application of these couplings to antikaon condensation in hypernuclear matter, examines the resulting matter properties, and discusses the implications for stellar structure in both cases. Finally, Section~\ref{sec:conclu} presents our conclusions. Throughout this work, we adopt natural units with $\hbar = c = 1$.

\section{Matter with RMF model}\label{sec:rmf}

We consider that the dense matter is composed of nucleons with electrons, and muons at low density regime, with the possibility of the appearance of antikaon condensates at high-density regime. This matter is modelled with the RMF model in which strong interaction between the hadrons is mediated by the isoscalar-scalar $\sigma$, isoscalar-vector $\omega^{\mu}$, $\phi^{\mu}$, and isovector-vector $\rho^{\mu}$ meson fields. Thus the total Lagrangian density is given by \cite{ 1999PhRvC..60b5803G, 2001PhRvC..64e5805B, 2018PhLB..783..234L, 2000NuPhA.674..553P}

\begin{equation}
\begin{aligned}
\mathcal{L} &= \sum_N \bar{\Psi}_{b}(i\gamma_{\mu}D^{\mu}_N-m^{*}_b)\Psi_{b}+\sum_{l=e,\mu} \bar{\Psi}_{l}(i\gamma_{\mu}\partial^{\mu}-m^{*}_l)\Psi_{l} + D^{\bar{K}*}_{\mu}\bar{K}D^{\mu}_{\bar{K}}K - m_{K}^{*^{2}}\bar{K}K 
+ \frac{1}{2}(\partial_{\mu}\sigma\partial^{\mu}\sigma - m_{\sigma}^{2}\sigma^{2}) \\
&\quad - \frac{1}{4}\omega_{\mu\nu}\omega^{\mu\nu} + \frac{1}{2}m^{2}_{\omega}\omega_{\mu}\omega^{\mu} 
- \frac{1}{4}\boldsymbol{\rho}_{\mu\nu} \cdot \boldsymbol{\rho}^{\mu\nu} + \frac{1}{2}m^{2}\boldsymbol{\rho}_{\mu} \cdot \boldsymbol{\rho}^{\mu} 
- \frac{1}{4}\phi_{\mu\nu}\phi^{\mu\nu} + \frac{1}{2}m^{2}_{\phi}\phi_{\mu}\phi^{\mu}.
\end{aligned}
\end{equation}
Here $N$ denotes nucleons, $\psi$ is the nucleonic wave function, $K$ is the kaonic wave function, $\sigma$ the $\sigma$-meson, $\omega_\mu$ the $\omega$-meson and $\rho_\mu$ the $\rho$-meson fields. The effective baryon and kaon masses are ${{m_b}^*}={m_b}-g_{\sigma{b}}\sigma$ and ${{m_K}^*}={m_K}-g_{\sigma{K}}\sigma$ respectively with $g_{\sigma j}$ the coupling of $\sigma$ meson with the hadron $j=N,K$. The antisymmetric field terms due to vector meson fields are given by 
\begin{eqnarray}
 \begin{aligned}   
\omega_{\mu \nu} = \partial_{\mu}\omega_{\nu} - \partial_{\nu}\omega_{\mu}, \\
\phi_{\mu \nu} = \partial_{\mu}\phi_{\nu} - \partial_{\nu}\phi_{\mu}, \\
\boldsymbol{\rho}_{\mu \nu} = \partial_{\mu} \boldsymbol{\rho}_{\nu} - \partial_{\nu}\boldsymbol{\rho}_{\mu}, \\
\end{aligned}
\end{eqnarray}
 and the covariant derivative is expressed as \cite{2001PhRvC..64e5805B}
\begin{equation}
D_{j\mu} = \partial_\mu + ig_{\omega j} \omega_\mu + ig_{\rho j} \boldsymbol{\tau}_{j3} \cdot \boldsymbol{\rho}_{\mu} + ig_{\phi j} \phi_\mu,
\end{equation}
where $j$ represents the nucleons $(N)$ and antikaons ($\bar{K}= K^{-}$ , $\bar{K}^{0}$ ). $g_{ij}$ are the couplings of the hadronic species $j$ with the mesons $i=\omega, \rho, \phi$. In the case of nucleons, we have the coupling $g_{\phi N} =0$ as the nucleons do not couple to the $\phi$ meson.

For the nucleonic sector, we consider the couplings from the density-dependent DDRH model. In this model, the isoscalar meson-nucleon couplings vary with density as
\cite{2020Parti...3..660T,2021PhRvD.103f3004T}
\begin{equation}\label{eqn.dd_isoscalar}
g_{i N}(n)= g_{i N}(n_{0}) f_i(x), \quad \quad \text{for }i=\sigma,\omega
\end{equation}
where the function is given by 
\begin{equation}\label{eqn.func}
f_i(x)= a_i \frac{1+b_i (x+d_i)^2}{1+c_i (x +d_i)^2},
\end{equation}
with $x=n/n_0$  where n is the total baryon number density and $n_0$ is the nuclear saturation density. The parameters $a_i$, $b_i$, $c_i$, and $d_i$ describe the density-dependent nature of the coupling parameters. The isovector-vector $\rho$-meson coupling with nucleons is given by \cite{2020Parti...3..660T,2021PhRvD.103f3004T} 
 
\begin{equation}
g_{\rho N}(n)= g_{\rho N}(n_{0}) e^{-a_{\rho}(x-1)}.
\end{equation}

In this model, the chemical potential for the nucleon is given as
\begin{eqnarray}
\mu_N = \sqrt{p_{F_N}^2 + m^{*2}_N} + \Sigma^0 + \Sigma^r,
\end{eqnarray}
where $p_{F_N}$ is the respective nucleon Fermi momentum.
Here, the vector self-energy is expressed as $\Sigma= \Sigma^0 + \Sigma^r $ with
\begin{eqnarray}
\Sigma^0 = g_{\omega b} \omega_0 + g_{\phi b} \phi_0 + g_{\rho b} \boldsymbol{\tau}_{b3} \rho_{03}
\end{eqnarray}
and $\Sigma^r$ is the rearrangement term. Here, the rearrangement term $\Sigma^{r}$ arises due to the density dependence of coupling parameters (to maintain thermodynamic consistency) and is given by 
\citep{2001PhRvC..64b5804H, 2002PhRvC..66f5801B, 2010PhRvC..81a5803T}  
\begin{equation}
\begin{aligned}
\Sigma^{r} & = \sum_{b} \left[ \frac{\partial g_{\omega b}}{\partial n}\omega_{0}n_{b} - \frac{\partial g_{\sigma b}}{\partial n} \sigma n_{b}^s + \frac{\partial g_{\rho b}}{\partial n} \rho_{03} \boldsymbol{\tau}_{b3} n_{b} + \frac{\partial g_{\phi b}}{\partial n}\phi_{0}n_{b} \right].
\end{aligned}
\end{equation} 
where $n=\sum_b n_b$ denotes the total baryon number density.

Strangeness-changing processes determine the threshold conditions for antikaons to appear, such as
$N \rightleftharpoons N + K$ and $e^- \rightleftharpoons K^{-}$ which are expressed by \cite{Glendenning:1997wn, 1997PhR...280....1P}
\begin{eqnarray}
\begin{aligned}
\mu_n - \mu_p = \omega_{K^-} = \mu_e , \quad \omega_{\bar{K}^0} = 0.
\end{aligned}
\end{eqnarray} 
where $\omega_{\bar{K}}$ is the in-medium energies of the antikaons given as (taking that for $K^-$, $\bar{K}^0$ isospin projection is $\mp \frac{1}{2}$)
\begin{eqnarray}
\omega_{K^-, \bar{K}^0} = m^{*}_K - g_{\omega K}\omega_0 - g_{\phi K} \phi_0 \mp g_{\rho K} \rho_{03}
\label{OmegabarK}
\end{eqnarray}
The number density of the s-wave antikaons is expressed as
\begin{equation}
\begin{aligned}
n_{K^-, \bar{K}^0} 
= - J_0^{K} = 2 \left(\omega_{\bar{K}} + g_{\omega K} \omega_0 + g_{\phi K} \phi_0 
   \mp \tfrac{1}{2} g_{\rho K} \rho_{03}\right) \bar{K} K = 2 m^{*}_K \bar{K} K
\end{aligned}
\end{equation}

where $J_\mu^K$ is the kaonic conserved current and is defined as,
$J_\mu^K = \bar{K}i\partial_\mu K - (i\partial_\mu\bar{K})K - 2g_{\omega K}\omega_\mu \bar{K}K - 2g_{\phi K}\phi_\mu \bar{K}K \mp \frac{1}{2} g_{\rho K}\rho_\mu \bar{K}K$.

The couplings in the nucleonic sector have been taken from the DDME2 parametrization  \cite{2020PhRvD.102l3007T}. For kaon couplings with mediator mesons, we calculate the couplings from SU(3) flavour symmetry, which is illustrated in the section \ref{sec:SU3}.

\subsection{Model parameters for nucleonic sector} 
Within this DDRH model, we use the parameters for the nucleonic sector from the parametrization DDME2 \cite{2020PhRvD.102l3007T}. For this parametrization, the parameter values in the nucleonic sector are tabulated in Table \ref{table:vector_coupling} \cite{2005PhRvC..71b4312L} with nucleon mass $m_N = 938.9$ MeV.

\begin{table*}[b]
	\begin{center}
 \caption{Values of the parameters for the DDME2 parametrization and Masses of mesons \cite{2005PhRvC..71b4312L}  at $n_0 = 0.152~ fm^{-3}$.}
 \resizebox{0.8\textwidth}{!}{%
\begin{tabular}{ccccccc}
     \hline
  Meson(i)  &$g_{iN}$ & $a_i$&$b_i$&$c_i$&$d_i$ & $m_i$(MeV)\\
   \hline
  $\sigma$  & 10.5396 & 1.3881 & 1.0943 &1.7057&0.4421& 550.1238 \\
  $\omega$  & 13.0189  & 1.3892 & 0.9240 & 1.4620&0.4775& 783 \\
  $\rho$  & 7.3672 & 0.5647 & -- & -- & -- & 763\\
   \hline
\end{tabular}
}%
\label{table:vector_coupling}
\end{center}
\end{table*}

\section{Antikaon couplings in SU(3) symmetry group}\label{sec:SU3}

The theory of strong interaction is invariant under the SU(3) flavour symmetry. In the case of strongly interacting meson particles, the invariant Yukawa Lagrangian can be constructed with interacting octet mesons and mediator nonet mesons. We consider the interacting mesons as antikaon mesons of the octet family ($J^P=0^-$), and the mediator mesons are isosinglet and isotriplet vector mesons of the nonet family ($J^P=1^-$). The Yukawa type interaction Lagrangian can be written as \cite{1963RvMP...35..916D, 2023PhRvD.107c6011L} 
\begin{equation}
    {\cal L}_{{\rm int}} = -g \bar{N}NM
\end{equation}
$N$ is the field for interacting mesons, and $M$ is the field for mediator vector meson $J^P=1^-$ family, including octet and singlet states. 

Interacting mesons in the pseudoscalar octet with $J^P = 0^-$ include the isospin doublets of kaons, $K \equiv (K^+, K^0)$ and antikaons, $\bar{K} \equiv (\bar{K}^0, K^-)$, the isospin triplet $\pi \equiv (\pi^+, \pi^0, \pi^-)$, and the isospin singlet $\eta_8$.  However, the physical $\eta$ and $\eta'$ mesons are not pure octet or singlet states; instead, they are admixtures of $\eta_8$ and the singlet $\eta_1$ \cite{PhysRev.135.B1076}, analogous to the way the physical $\omega$ and $\phi$ mesons arise from the mixing of $\omega_8$ and $\phi_1$. Based on two-photon decay data, Pham (1990)~\cite{Pham:1990tv} determined a mixing angle $\theta = -(18.4 \pm 2)^{\circ}$, which reconciles the theoretical prediction for the $\eta \to \gamma\gamma$ decay width experiment. In this mixed-state basis, the physical states are expressed as  
\begin{equation}
\eta = \cos\theta \,\eta_8 - \sin\theta \,\eta_1, 
\qquad 
\eta' = \sin\theta \,\eta_8 + \cos\theta \,\eta_1 .
\end{equation}

There are three parameters in the flavour SU(3) symmetry: the weight factor $\alpha_v$, the ratio $z = g_8 / g_1$, and the mixing angle $\theta_v$. The weight factor for the contributions of the symmetric D and antisymmetric F couplings about one another is $\alpha_v = F/(F+D)$, by definition, limited to the interval $ 0 \leq \alpha_v \leq 1$, where a pure D-type coupling corresponds to the lower bound and a pure F-type coupling corresponds to the upper limit \cite{2012PhRvC..85f5802W, 1979PhRvD..20.1633N, 1999PhRvC..59...21R}. The relative strength of the coupling with the meson octet ($g_8$) over the singlet ($g_1$) is represented by the ratio z. The z and $\alpha_v$ parameters represent the relationship between the coupling's nature and relative strength \cite{2023PhRvD.107c6011L}. Clebsch--Gordon coefficients are needed to compute couplings that are functions of the free parameter 
$\alpha_v$. The nature of $\omega$ and $\phi$ meson is expressed by the mixing angle $\theta_v$ \cite{1979PhRvD..20.1633N, 1999PhRvC..59...21R}.\\

We introduce the couplings as given by \cite{1963RvMP...35..916D}
\begin{equation}
    g_{8}=\frac{\sqrt{30}}{40}g_{s}+\frac{\sqrt{6}}{24}g_{a}
    ~~~ {\rm and} ~~~ \alpha_v=\frac{\sqrt{6}}{24}\frac{g_{a}}{g_{8}}.
\end{equation} 
The constants corresponding to the antisymmetric and symmetric coupling are given as $g_{a}$ and $g_{s}$, respectively.

Thus, we can rewrite the couplings in terms of these parameters for $M$ belonging to the octet state as, 
\begin{eqnarray}
g_{\omega_{8}K}=\frac{1}{3}g_8\sqrt{3}(4\alpha_{v}-1)\\
g_{\omega_{8}\bar{K}}=-\frac{1}{3}g_8\sqrt{3}(1+2\alpha_{v})\\
g_{\omega_{8}\pi}=\frac{2}{3}g_8\sqrt{3}(1-\alpha_{v})\\
g_{\omega_{8}\eta_8}=-\frac{2}{3}g_8\sqrt{3}(1-\alpha_{v})\\
g_{\rho K}=g_{8}\\
g_{\rho \bar{K}}= -g_{8}(1-2\alpha_v) \label{kbarrho}   \\
g_{\rho \pi}=  2g_{8}\alpha_v \label{pirho}\\
g_{\rho \eta_8}=0
\end{eqnarray}
and
\begin{equation}
g_{\phi_{1}K} = g_{\phi_{1}\bar{K}} = g_{\phi_{1}\pi} = g_{\phi_{1}\eta_8} = g_1
\end{equation}

In nature, the physical realisation of the isospin singlet vector mesons are $\omega$ and $\phi$ mesons, which are a mixture of the theoretical $\omega_{8}$ and $\phi_{1}$ states like \cite{1984PrPNP..12..171D}
\begin{eqnarray}
    \omega= \cos \theta_{v}\ket{\phi_{1}} + \sin \theta_{v}\ket{\omega_{8}} \\
    \phi=-\sin \theta_{v}\ket {\phi_{1}}+ \cos \theta_{v}\ket{\omega_{8}},
\end{eqnarray}
where $\theta_v$ is the mixing angle between the states $\omega_8$ and $\phi_1$. Therefore, the couplings of the physical $\omega$ meson are
\begin{eqnarray}
			g_{\omega K} 
			= g_1\cos \theta_v + g_8 \sin \theta_v \frac 
			1{\sqrt{3}} (4\alpha_v -1) \\
			g_{\omega \bar{K}} 
			= g_1\cos \theta_v - g_8 \sin \theta_v \frac 
			1{\sqrt{3}} (1+2\alpha_v) \label{kbaromega} \\
			g_{\omega \pi} 
			= g_1\cos \theta_v + g_8 \sin \theta_v \frac 
			2{\sqrt{3}} (1-\alpha_v) \\
			g_{\omega \eta_8} 
			= g_1\cos \theta_v - g_8 \sin \theta_v \frac 
			2{\sqrt{3}} (1-\alpha_v).    
\end{eqnarray}
The couplings with $\phi$ mesons can be obtained just by substituting $\sin\theta_v$ and $\cos\theta_v$ of the expressions of couplings for $\omega$ mesons by $\cos\theta_v$ and $-\sin\theta_v$ respectively \cite{1984PrPNP..12..171D}.\\

\subsection{Constrained on the parameters}
For the ``ideal mixing" for the $\omega$ and $\phi$ meson we get 
\begin{equation}
    \tan {\theta_v}= \frac{1}{\sqrt{2}}
    \label{tanthetav}
\end{equation}
The ideal mixing angle $\theta_v\approx 35.3^\circ$ is rather close to $\theta_v \approx 40^\circ $, which is obtained using the quadratic mass formula for the mesons. Therefore, we maintain the ideal mixing condition for the vector mesons \cite{2012NuPhA.881...62W}. And the range of $0\le \alpha_v \le 1$ from the definition as explained above. 

Using the Eq. \eqref{tanthetav}
\begin{equation}
    \frac{g_{\omega\bar{K}}}{g_{\omega N}}= \frac{\sqrt{6}-z_m(1+2\alpha_m)}{\sqrt{6}+z_b(4\alpha_b-1)}
    \label{ratio_omegabarK}
\end{equation}
where \cite{2023PhRvD.107c6011L}
\begin{equation}
   g_{\omega N}=g_1'\cos \theta_v + g_8' \sin \theta_v \frac1{\sqrt{3}} (4\alpha_b -1). 
   \label{bary}
\end{equation}
The mesonic and baryonic octets originate from distinct SU(3) representations, with the mesonic octet from quark–antiquark states and the baryonic octet from three-quark states. Consequently, their corresponding couplings are introduced independently, and no inherent correlation exists between them within the present framework. In this framework, $\alpha_v$ and $z$ are represented by $\alpha_m$ and $z_m$ in the mesonic sector, and by $\alpha_b$ and $z_b$ in the baryonic sector. In the Eq. \eqref{bary} $g_1'$ and $g_8'$ correspond to the baryonic sector singlet and octet couplings. \\

We require the $\omega$ couplings to be attractive for kaons. In contrast, for nucleons ($p, n$) the $\omega$ coupling contributes repulsively, whereas for kaons the vector interaction enters with the opposite sign, making the $\omega$ contribution attractive. Therefore, consistency with Eq.~\eqref{OmegabarK} \cite{1999PhRvC..60b5803G} demands that $g_{\omega \bar{K}}$ be positive in all cases.
\begin{equation}
\frac{g_{\omega\bar{K}}}{g_{\omega N}} \geq 0
\end{equation}  
This requires the range of {$z_m$ to be $0 \le z_m \le 2/ \sqrt{6}$ from the possibility that $\alpha_v$ can only have the range $0 \le \alpha_v \le 1$.

We can fix either the $z_m$ value and vary $\alpha_m$ or fix the $\alpha_m$ and vary the $z_m$ value as has been done for SU(3) hyperon couplings by Weissenborn {\it{et al.}} \cite{2012PhRvC..85f5802W}. 

\subsection{Fixing the $z_m$ value from SU(6) symmetry}

Requirement of spin-independence for the $qq\omega$ and $qq\phi$ couplings within the identically flavoured $K$ and $\bar{K}$ gives
\begin{eqnarray}
	g_{K\omega} - g_{\bar{K}\omega} = 0
	\nonumber \\
	{\rm or} ~~~ g_8\sin\theta_v\frac 1{\sqrt{3}} (4\alpha_v-1+1+2\alpha_v) = 0
	\nonumber \\
	{\rm or} ~~~ g_8\sin\theta_v\frac 1{\sqrt{3}} 6\alpha_v = 0.
\end{eqnarray}
From here, we fix $\alpha_v = 0$ as $g_8$ and $\theta_v$ can not be zero.

From quark composition for mesonic sector requirement of $g_{\pi\phi} = 0$ leads to
\begin{eqnarray}
	g_{\pi\phi} = -g_1\sin \theta_v + g_8 \cos \theta_v \frac 
			2{\sqrt{3}} = 0
	\nonumber \\
	{\rm or}~~~ \frac {g_1}{g_8} \tan\theta_v = \frac 2{\sqrt{3}}.
     \label{eq1}
 \end{eqnarray}
By giving the ideal mixing value of $\theta_v$ \eqref{tanthetav} we get the value of $z$ as
\begin{equation}
    z=\frac{\sqrt{3}}{2\sqrt{2}}
\end{equation}

 From this, with $\alpha_v=0$, we get a value of $z$ from SU(6) symmetry. \\

In the exact SU(6) limit, the couplings reduce to the simple relations
\( g_{\omega K} = g_{\omega \bar{K}} \) and \( g_{\phi \pi} = 0 \). 
Once SU(6) symmetry is relaxed, however, these relations no longer hold. 
In the more general SU(3) framework since \(K\) and \(\bar{K}\) 
differ in both hypercharge (\(Y\)) and isospin (\(I\)), their octet couplings involve different Clebsch--Gordan coefficients, which naturally leads to 
\( g_{\omega K} \neq g_{\omega \bar{K}} \). A detailed derivation of such SU(3) 
Clebsch--Gordan coefficients for baryon--meson couplings are given in 
\cite{PhysRevD.70.096015}. 
Thus, the difference between particle and antiparticle couplings arises directly 
from the group-theoretical structure of SU(3). Phenomenology provides further support for this picture. Relativistic heavy-ion 
collision data show that particles and antiparticles display different elliptic 
flows, pointing to distinct in-medium interactions 
\cite{Xu:2014qra}. 
Hence, both symmetry arguments and experimental evidence consistently 
indicate that particle--antiparticle couplings can differ in realistic environments. \\

If we fix $z_m$ to its SU(6) value $z_m=\sqrt{3}/2\sqrt{2}$ and use ideal mixing while varying $\alpha_m$ between $0\le\alpha_m \le1$ we obtain vector couplings for three different cases. Since the parameter values of ($\alpha_b$, $z_b$) in the baryonic sector and ($\alpha_m$, $z_m$) in the mesonic sector are independent and are not interrelated, we fix $z_b=z_m$, and for $\alpha_b$,  we consider three possible scenarios. In the first case, we consider $\alpha_b=\alpha_m$, and the corresponding couplings of octet mesons are given in Table \ref{table:vector_coupling_meson}. The second case corresponds to $\alpha_b = 2\alpha_m$ for which the meson couplings are shown in Table \ref{table:vector_coupling_meson2}. Here, we determine the couplings for $\alpha_m$ up to a maximum value of $0.5$, since taking $\alpha_b = 2\alpha_m$ implies that any $\alpha_m > 0.5$ would correspond to $\alpha_b > 1$, which is not allowed because, by definition, $0 \le \alpha_v \le 1$ \cite{2012PhRvC..85f5802W, 1979PhRvD..20.1633N, 1999PhRvC..59...21R}. The third case takes $\alpha_b=\alpha_m/2$ for which the meson couplings are given in Table \ref{table:vector_coupling_meson3}.\\

We have also included in the coupling tables the corresponding values obtained from the simple QMIC rule. According to this rule, the coupling of an interacting meson with the $\omega$ mediator meson depends on the number of non-strange quarks it contains. For example, in the case of kaons, we have  
\[
\frac{g_{\omega K}}{g_{\omega N}} = \frac{1}{3},
\]  
Since a kaon contains one non-strange quark, whereas nucleons contain three.  

For the $\rho$ meson as the mediator, the coupling is determined by the total isospin. Therefore,  
\[
\frac{g_{\rho K}}{g_{\rho N}} = 1,
\]  
because both kaons and nucleons have total isospin $I=\tfrac{1}{2}$ \cite{1994AnPhy.235...35S,2001PhRvC..64e5805B,2021ApJ...910...96M,  1963RvMP...35..916D,  2011PhRvC..84c5809R, 2014PhRvC..90a5801C}.  

For the $\phi$ meson coupling, the value of $g_{\phi K}$ is obtained from the relation involving the $\rho \rightarrow 2\pi$ decay width \cite{PhysRevC.80.035205, 2011JPhCS.312b2013G}:  
\[
\sqrt{2}\, g_{\phi K} = g_{\rho \pi} = 6.04.
\]  

Finally, in the case of pions ($\pi$) and the octet eta ($\eta_8$), there are no strange quarks present. Hence, they cannot couple with the $\phi (s\bar{s})$ meson.

\begin{table*}  

\begin{center}
 \caption{Numerical values of vector couplings for different values of $\alpha_m$ with $\alpha_b= \alpha_m$ }.
 \resizebox{1.1\textwidth}{!}{%
 \begin{tabular}{ccccccccccccc}
  \hline
 $\alpha_m$& 0.0 & 0.1 & 0.2 & 0.3 & 0.4 & 0.5 & 0.6 & 0.7 & 0.8 & 0.9 & 1.0&QMIC \\
  \hline
  $g_{\omega K}$ & 13.02& 13.02& 13.02&13.02&13.02&13.02& 13.02& 13.02& 13.02& 13.02& 13.02&4.33\\
$g_{\rho K}$   & 7.37 & 7.37  & 7.37 & 7.37 & 7.37 &7.37 & 7.37& 7.37 & 7.37  &7.37  &7.37 &7.36\\
$g_{\phi K}$   & -18.41 & -14.08& -10.66 & -7.89 & -5.60 & -3.68 & -2.05 & -0.63 &0.59& 1.67 & 2.63 &-4.27\\
\hline
$g_{\omega \bar{K}}$ & 13.02 & 10.72 & 8.91 & 7.44& 6.23 & 5.21 & 4.34 & 3.59 & 2.94 & 2.37 & 1.86 &4.33 \\
$g_{\rho \bar{K}}$   & -7.37 & -5.89 & -4.42 & -2.95 & -1.47 & 0.00 & 1.47 & 2.95 & 4.42 & 5.89 & 7.37&7.36 \\
$g_{\phi \bar{K}}$   & -18.41 & -17.33 & -16.47 & -15.78 & -15.21 & -14.73 & -14.32 & -13.97 & -13.66 & -13.39 & -13.15 &-4.27\\
\hline
$g_{\omega \pi}$ & 26.04 & 22.21 & 19.19 & 16.74 & 14.72 & 13.02 & 11.57 & 10.33 & 9.24 & 8.28 & 7.44&8.68 \\
$g_{\rho \pi}$   & 0.00  & 1.47  & 2.95  & 4.42  & 5.89  & 7.37 & 8.84 & 10.31 & 11.79&13.26 &  14.73 &14.72\\
$g_{\phi \pi}$   & 0.00  & -1.08 & -1.94& -2.63 & -3.20 & -3.68& -4.09 & -4.44 & -4.75 & -5.02 & -5.26&0.00 \\
\hline
$g_{\omega \eta_8}$ & 8.68  & 8.42  & 8.22  & 8.06  & 7.92  & 7.81  & 7.71  & 7.63  & 7.56  & 7.50  & 7.44  & 8.68\\
$g_{\rho \eta_8}$   & 0.00  & 0.00  & 0.00  & 0.00  & 0.00 & 0.00  & 0.00  & 0.00  & 0.00  & 0.00  & 0.00  & 0.00\\
$g_{\phi \eta_8}$   & -24.55& -20.58& -17.44& -14.90& -12.81& -11.05& -9.55 & -8.25 & -7.13 & -6.14 & -5.26& 0.00\\
\hline

\end{tabular}
}%
\label{table:vector_coupling_meson}
\end{center}

\begin{center}
 \caption{Numerical values of vector couplings for different values of $\alpha_m$ with $\alpha_b= 2\alpha_m$.}
 \resizebox{0.65\textwidth}{!}{%
 \begin{tabular}{cccccccc}
  \hline
 $\alpha_m$& 0.0 & 0.1 & 0.2 & 0.3 & 0.4 & 0.5 & QMIC \\
  \hline
  $g_{\omega K}$ & 13.02 & 13.02 & 13.02 & 13.02 & 13.02& 13.02 &4.33 \\
$g_{\rho K}$   & 7.37  & 7.37  & 7.37  & 7.37  & 7.37& 7.37  &7.36\\
$g_{\phi K}$   & -18.41 & -12.60 & -8.81 & -6.14 &-4.16& -2.63 &-4.27\\
\hline
$g_{\omega \bar{K}}$ & 13.02 & 9.59 & 7.36 & 5.79 & 4.62 & 3.72 &4.33 \\
$g_{\rho \bar{K}}$   & -7.37 & -5.89 & -4.42 & -2.95 & -1.47 & 0.00 &7.36 \\
$g_{\phi \bar{K}}$   & -18.41 & -15.50 & -13.61 & -12.27 & -11.28&-10.52 &-4.27\\
\hline
$g_{\omega \pi}$ & 26.04 & 19.87 & 15.85 &13.02& 10.92 & 9.30 &8.68 \\
$g_{\rho \pi}$   & 0.00  & 1.47  & 2.95  & 4.42  & 5.89  &7.37&14.72\\
$g_{\phi \pi}$   & 0.00  & -0.97 & -1.60 & -2.05 & -2.38&-2.63 &0.00 \\
\hline
$g_{\omega \eta_8}$ & 8.68 & 7.54 & 6.79 &6.27& 5.88 & 5.58 &8.68\\
$g_{\rho \eta_8}$   & 0.00 & 0.00 & 0.00 & 0.00 & 0.00 & 0.00 & 0.00\\
$g_{\phi \eta_8}$   & -24.55 & -18.41 & -14.41 & -11.59 &-9.50& -7.89&0.00 \\
\hline
 
\end{tabular}
}%
\label{table:vector_coupling_meson2}
\end{center}

\begin{center}
 \caption{Numerical values of vector couplings for different $\alpha_m$, with $\alpha_b = \alpha_m/2$.}
 \resizebox{1.1\textwidth}{!}{%
 \begin{tabular}{ccccccccccccc}
  \hline
 $\alpha_m$& 0.0 & 0.1 & 0.2 & 0.3 & 0.4 & 0.5 & 0.6 & 0.7 & 0.8 & 0.9 & 1.0&QMIC \\
  \hline
  $g_{\omega K}$ & 13.02 & 13.02 & 13.02 & 13.02 & 13.02 & 13.02 & 13.02 & 13.02 & 13.02 & 13.02 & 13.02&4.33 \\
$g_{\rho K}$   & 7.37  & 7.37  & 7.37  & 7.37  & 7.37  & 7.37  & 7.37  & 7.37  & 7.37  & 7.37  & 7.37 &7.36\\
$g_{\phi K}$   & -18.41 & -14.96 & -11.91 & -9.21 & -6.78 & -4.60 & -2.63 & -0.84 & 0.80 & 2.30 & 3.68 &-4.27\\
\hline
$g_{\omega \bar{K}}$ & 13.02 & 11.39 & 9.96 & 8.68 & 7.54 & 6.51 & 5.58 & 4.73 & 3.96 & 3.25 & 2.60 &4.33 \\
$g_{\rho \bar{K}}$   & -7.37 & -5.89 & -4.42 & -2.95 & -1.47 & 0.00 & 1.47 & 2.95 & 4.42 & 5.89 & 7.37&7.36 \\
$g_{\phi \bar{K}}$   & -18.41 & -18.41 & -18.41 & -18.41 & -18.41 & -18.41 & -18.41 & -18.41 & -18.41 & -18.41 & -18.41 &-4.27\\
\hline
$g_{\omega \pi}$ & 26.04 & 23.60 & 21.44 & 19.53 & 17.82 & 16.27 & 14.88 & 13.61 & 12.45 & 11.39 & 10.42&8.68 \\
$g_{\rho \pi}$   & 0.00  & 1.47  & 2.95  & 4.42  & 5.89  & 7.37  & 8.84  & 10.31 & 11.79 & 13.26 & 14.73 &14.72\\
$g_{\phi \pi}$   & 0.00  & -1.15 & -2.17 & -3.07 & -3.88 & -4.60 & -5.26 & -5.86 & -6.40 & -6.90 & -7.36&0.00 \\
\hline
$g_{\omega \eta_8}$ & 8.68 & 8.95 & 9.19 & 9.40 & 9.59 & 9.76 & 9.92 & 10.06 & 10.19 & 10.31 & 10.42 &8.68\\
$g_{\rho \eta_8}$   & 0.00 & 0.00 & 0.00 & 0.00 & 0.00 & 0.00 & 0.00 & 0.00 & 0.00 & 0.00 & 0.00 &0.00\\
$g_{\phi \eta_8}$   & -24.55 & -21.86 & -19.49 & -17.39 & -15.50 & -13.81 & -12.27 & -10.88 & -9.61 & -8.44 & -7.36&0.00 \\
\hline

\end{tabular}
}%
\label{table:vector_coupling_meson3}
\end{center}

\end{table*}

 The scalar meson coupling parameter is calculated from the relation of the real component of $K^{-}$ optical potential depth at nuclear saturation density \cite{2014PhRvC..89b5805L, 2012PhRvC..85f5802W}.

In nuclear matter, antikaons possess an attractive potential \cite{2019Parti...2..411M, 1997NuPhA.625..287W, 1999PhRvC..60b4314F,1997NuPhA.625..372L,2000PhRvC..62f1903P}, while kaons have the opposite effect \cite{1997NuPhA.625..372L, 2000PhRvC..62f1903P}.
Corresponding to different values of  $\alpha_m$, the values of antikaon-scalar meson coupling parameters are tabulated in Table \ref{table:gsigmak}.
The value of the antikaon optical potential depth ($U_{\bar{K}}$) in nuclear matter has been estimated using several theoretical and experimental approaches, leading to a wide range of possible values. The value of $K^{-}$ optical potential depth in various literatures is reported to be in the range from -40 to -200 MeV \cite{1997NuPhA.625..287W, 1999PhRvC..60b4314F, Koch:1994mj, LUTZ199812, SCHAFFNERBIELICH2000153}. One method involves the analysis of low-energy scattering data between $K^-$ mesons and protons, which suggests a relatively shallow potential in the range of approximately $-120 \leq U_{\bar{K}} \leq -40$ MeV \cite{Koch:1994mj}. A second approach fits relativistic mean field (RMF) potentials to kaonic atom data, yielding significantly deeper potentials around $U_{\bar{K}} = -180 \pm 20$ MeV \cite{1999PhRvC..60b4314F}. Additionally, more moderate values in the range of $-80 \leq U_{\bar{K}} \leq -50$ MeV are obtained from self-consistent calculations based on chiral Lagrangians or meson-exchange models, which incorporate many-body effects and medium modifications more systematically \cite{SCHAFFNERBIELICH2000153, Mishra:2008dj, PhysRevC.65.054907}. These differing predictions reflect the ongoing uncertainty in constraining $U_{\bar{K}}$.
Recently, using Bayesian analysis with constraints from $\chi$EFT calculations, nuclear saturation properties, and astrophysical observations, the value of \( U_{\bar{K}} \) was estimated to be \( -129.36^{+12.53}_{-3.837} \) MeV on a 68 \% confidence interval \cite{2024PhRvC.110d5804P}. Consequently, in our current calculations, we choose the value of \( U_{\bar{K}} = -130 \) MeV within this range.
Table \ref{table:gsigmak} shows the values of $g_{\sigma\bar{K}}$ for different values of  $\alpha_m$ for $U_{\bar{K}} = -130$ MeV for the three seperate cases of $\alpha_b = \alpha_m$, $ 2\alpha_m$ and $\alpha_m$/2.

\begin{table*} 

\begin{center} 
 \caption{The antikaon-scalar meson coupling parameter
values in SU(3) for different $\alpha_m$ and QMIC \cite{2021PhRvD.103f3004T} case for $U_{\bar{K}} = -130$ MeV at $n_0 = 0.152~fm^{-3}$, shown for the three cases.}

\resizebox{1.12\textwidth}{!}{%
\begin{tabular}{cccccccccccccc}
  \hline
   $\alpha_m$&0&0.1&0.2&0.3& 0.4 & 0.5 &0.6 & 0.7 & 0.8 & 0.9 & 1 & QMIC\\
  \hline
  $g_{\sigma\bar{K}}(\alpha_b=\alpha_m)$&-4.9491 &-3.4554&-2.2764&-1.3217&-0.5330&0.1288&0.6933&1.1799& 1.6031&1.9755 & 2.3050& 0.6930 \\
  $g_{\sigma\bar{K}}(\alpha_b=2\alpha_m)$&-4.9491 &-2.7216&-1.2693&-0.2469&0.5113&1.0961&1.5616&1.9399& 2.2539&2.5188 & 2.7448& 0.6930 \\
  $g_{\sigma\bar{K}}(\alpha_b=\alpha_m/2)$&-4.9491&-3.8912&-2.9577&-2.1275&-1.3852&-0.7175&-0.1123&0.4368&0.9387 & 1.3983&1.8215 &0.6930 \\
   \hline  
\end{tabular}
}
\label{table:gsigmak}
\end{center}

\end{table*}

\subsection{Fixing the  $\alpha_m$ value}

In this section, we fix $\alpha_m$ and vary the $z_m$ values. Similar to the analysis with $\alpha_m$, we consider three different cases here as well, fixing $\alpha_b=\alpha_m$. 

In the first case, we take $z_b = z_m$.} We then fix $\alpha_m$ to a representative value, $\alpha_m=0.6$, and vary the parameter $z_m$ within its allowed range, $0 \le z_m \le 2/\sqrt{6}$. We do not choose $\alpha_m=0$, its SU(6) value, since this would lead to the ratio in Eq.~\eqref{ratio_omegabarK} becoming unity, which prevents us from varying $z_m$ within its range and studying its influence on the $\omega$–antikaon coupling.  

For the second and third cases, with $z_b = 2z_m$ and $z_b = z_m/2$, respectively, we fix $\alpha_m$ to its SU(6) value, and determine the corresponding couplings. Using the chosen parameter values, we calculate the corresponding vector meson couplings for the cases of $z_b=z_m$, $2z_m$ and $z_m/2$ as presented in the Tables \ref{table:vector_coupling_meson_z}, \ref{table:vector_coupling_meson_2z} and \ref{table:vector_coupling_meson_z/2}, respectively. In Table \ref{table:vector_coupling_meson_2z}, the values of $z_m$ considered range from 0 to 0.4. This is because we impose the condition $z_b = 2z_m$, so any $z_m > 0.4$ would correspond to $z_b > 0.8$, which is not allowed due to the constraint $0 \leq z_m \leq 2/\sqrt{6}$. \\

\begin{table*}  

	\begin{center}
 \caption{Numerical values of vector couplings for different values of $z_m$ with $z_b = z_m$.}
 \resizebox{0.9\textwidth}{!}{%
 \begin{tabular}{ccccccccccc}
    \hline
  $z_m$ & $0$&$0.1$&$0.2$&$0.3$&$0.4$&$0.5$&$0.6$&$0.7$&$0.8$&QMIC \\
 \hline
  $g_{\omega K}$ & 13.02 & 13.02 & 13.02 & 13.02 & 13.02 & 13.02 & 13.02 & 13.02 & 13.02&4.33 \\
$g_{\rho K}$   & 7.37& 7.37  & 7.37  & 7.37  & 7.37  & 7.37  & 7.37 & 7.37  & 7.37  &7.36\\
$g_{\phi K}$   & -9.21 & -7.71 & -6.37 & -5.16 & -4.07 & -3.07  & -2.15  &-1.31  & -0.54  &-4.27\\
\hline
  $g_{\omega \bar{K}}$ & 13.02 & 11.21 & 9.58 & 8.12 & 6.79 & 5.58 &4.47&3.45&2.51&4.33 \\ 
  $g_{\rho\bar{K}}$ & 1.47&1.47 & 1.47 &1.47&1.47& 1.47 & 1.47&1.47&1.47&7.36 \\
  $g_{\phi\bar{K}}$&-9.21&-10.27&-11.23&-12.09&-12.88 &-13.59&-14.24&-14.84&-15.40&-4.27 \\
\hline
$g_{\omega \pi}$ & 13.02 & 12.72 & 12.45 & 12.20 & 11.98 & 11.78 & 11.59 & 11.42 & 11.27 & 8.68 \\
$g_{\rho \pi}$   & 8.84  & 8.84  & 8.84  & 8.84 & 8.84 & 8.84 & 8.84  & 8.84 &8.84 & 14.72\\
$g_{\phi \pi}$   & -9.21  & -8.14 & -7.18 & -6.32 & -5.54 & -4.82 & -4.17 & -3.57 & -3.02 & 0.00\\
\hline
$g_{\omega \eta_8}$ &13.02 & 11.91 & 10.92 & 10.02 & 9.21 & 8.47 & 7.79 & 7.17 & 6.60 &8.68\\
$g_{\rho \eta_8}$   & 0.00 & 0.00 & 0.00 & 0.00 &  0.00 &  0.00 &  0.00 &  0.00 &  0.00 &0.00\\
$g_{\phi \eta_8}$   & -9.21 & -9.28 & -9.34 & -9.40 & -9.45 & -9.50 & -9.54 & -9.58& -9.62 & 0.00 \\
\hline
  
\end{tabular}
}%
\label{table:vector_coupling_meson_z}
\end{center}

\begin{center}
 \caption{Numerical values of vector couplings for different values of $z_m$ with $z_b= 2z_m$.}
 \resizebox{0.6\textwidth}{!}{%
 \begin{tabular}{cccccccc}
  \hline
 $z_m$& 0.0 & 0.1 & 0.2 & 0.3 & 0.4  &QMIC \\
  \hline
  $g_{\omega K}$ & 13.02 & 13.02 & 13.02 & 13.02 & 13.02&4.33 \\
$g_{\rho K}$   & 7.37  & 7.37  & 7.37  & 7.37  & 7.37&7.36\\
$g_{\phi K}$   & -9.21 & -7.32 & -5.78 & -4.50 &-3.43& -4.27\\
\hline
$g_{\omega \bar{K}}$ & 13.02 & 10.63 & 8.69 & 7.08 & 5.72 &4.33 \\
$g_{\rho \bar{K}}$   & 1.47 & 1.47 & 1.47 & 1.47 & 1.47 & 7.36 \\
$g_{\phi \bar{K}}$   & -9.21 & -9.75 & -10.18 & -10.55 & -10.86&-4.27\\
\hline
$g_{\omega \pi}$ & 13.02 & 12.06 & 11.29 &10.64& 10.10 &8.68 \\
$g_{\rho \pi}$   & 8.84  & 8.84  & 8.84  & 8.84  & 8.84  &14.72\\
$g_{\phi \pi}$   & -9.21  & -7.72 & -6.51 & -5.51 & -4.67&0.00 \\
\hline
$g_{\omega \eta_8}$ & 13.02 & 11.30 & 9.90 &8.74& 7.77 &8.68\\
$g_{\rho \eta_8}$   & 0.00 & 0.00 & 0.00 & 0.00 & 0.00 & 0.00\\
$g_{\phi \eta_8}$   & -9.21 & -8.80 & -8.47 & -8.20 &-7.97&0.00 \\
\hline
 
\end{tabular}
}%
\label{table:vector_coupling_meson_2z}
\end{center}

\begin{center}
 \caption{Numerical values of vector couplings for different $z_m$, with $z_b = z_m/2$.}
 \resizebox{0.9\textwidth}{!}{%
 \begin{tabular}{ccccccccccc}
  \hline
 $z_m$& 0.0 & 0.1 & 0.2 & 0.3 & 0.4 & 0.5 & 0.6 & 0.7 & 0.8 &QMIC \\
  \hline
  $g_{\omega K}$ & 13.02 & 13.02 & 13.02 & 13.02 & 13.02 & 13.02 & 13.02 & 13.02 & 13.02 &4.33 \\
$g_{\rho K}$   & 7.37  & 7.37  & 7.37  & 7.37  & 7.37  & 7.37  & 7.37  & 7.37  & 7.37 &7.36\\
$g_{\phi K}$   & -9.21 & -7.93 & -6.72 & -5.57& -4.48 & -3.45 & -2.47 & -1.53 & -0.64 &-4.27\\
\hline
$g_{\omega \bar{K}}$ & 13.02 & 11.52 & 10.10& 8.76 & 7.49 & 6.28 & 5.58 & 5.12 & 4.03 & 4.33 \\
$g_{\rho \bar{K}}$   & 1.47 & 1.47 & 1.47 & 1.47 & 1.47 & 1.47 & 1.47 & 1.47 & 1.47 &7.36 \\
$g_{\phi \bar{K}}$   & -9.21 & -10.56 & -11.84 & -13.05 & -14.20 & -15.29 & -16.33 & -17.32 & -18.26 &-4.27\\
\hline
$g_{\omega \pi}$ & 13.02 & 13.07 & 13.12 & 13.17 & 13.21 & 13.25 & 13.29 & 13.33 & 13.36 &8.68 \\
$g_{\rho \pi}$   & 8.84  & 8.84  & 8.84  & 8.84  & 8.84  & 8.84  & 8.84  & 8.84 & 8.84 &14.72\\
$g_{\phi \pi}$   & -9.21  & -8.37 & -7.57 & -6.82 & -6.10 & -5.42 & -4.78 & -4.16 & -3.58 &0.00 \\
\hline
$g_{\omega \eta_8}$ & 13.02 & 12.24 & 11.51 & 10.82 & 10.16 & 9.53 & 8.94 & 8.37 & 7.83 &8.68\\
$g_{\rho \eta_8}$   & 0.00 & 0.00 & 0.00 & 0.00 & 0.00 & 0.00 & 0.00 & 0.00 & 0.00 &0.00\\
$g_{\phi \eta_8}$   & -9.21 & -9.53 & -9.85 & -10.14 & -10.42 & -10.69 & -10.94 & -11.18 & -11.41&0.00 \\
\hline

\end{tabular}
}%
\label{table:vector_coupling_meson_z/2}
\end{center}

\end{table*}

\begin{table} 

	\begin{center} 
 \caption{The antikaon-scalar meson coupling parameter
values in SU(3) for different $z_m$ \cite{2021PhRvD.103f3004T} case for $U_{\bar{K}} = -130$ MeV at $n_0 = 0.152~fm^{-3}$, shown for the three cases.}
\resizebox{1.0\textwidth}{!}{%
\begin{tabular}{cccccccccccc}
  \hline
   $z_m$& $0$ & $0.1$ &$0.2$ & $0.3$ & $0.4$ & $0.5$ & $0.6$ & $0.7$ &$0.8$&QMIC\\
  \hline
  $g_{\sigma\bar{K}}(z_b=z_m)$&-4.9491&-3.7643&-2.7487&-1.7331& -0.8867&-0.1250 & 0.6366&1.2290&1.9061& 0.6930 \\
  $g_{\sigma\bar{K}}(z_b=2z_m)$&-4.9491&-5.2877&-5.7955&-6.3033& -6.9803&--&--&--&--& 0.6930 \\
  $g_{\sigma\bar{K}}(z_b=z_m/2)$&-4.9491&-4.7799&-4.6106&-4.3567&-4.1874&-4.0182 &-3.7643&-3.5104&-3.2565& 0.6930 \\

   \hline  
\end{tabular}
}
\label{table:gsigmak_z}
\end{center}

\end{table}

For all the cases, where we take a fixed $\alpha_m$ value and vary $z_m$, as represented in Tables \ref{table:vector_coupling_meson_z}, \ref{table:vector_coupling_meson_2z}, and \ref{table:vector_coupling_meson_z/2}, the $\omega$ couplings for antikaons decrease, while the $\phi$ couplings increase with increasing $z_m$ values. However, the $\rho$ couplings remain constant, as the value of $\alpha_m$ is fixed, since the ratios of the $\rho$ couplings to $g_{\omega N}$ depend only on $\alpha_m$.

The scalar meson coupling constants in the present analysis are determined following the same approach as described previously, wherein the coupling is extracted from the real part of the $K^{-}$ optical potential depth $U_{\bar{K}}$ at nuclear saturation density. Specifically, the values of $g_{\sigma \bar{K}}$ are obtained for $U_{\bar{K}} = -130$ MeV using this relation, and are listed in Table \ref{table:gsigmak_z} for the three cases of $z_b = z_m$, $2z_m$ and $z_m/2$.

\section{Application to dense matter with antikaon condensate}\label{sec:appli}
With the calculated kaon couplings for SU(3) flavour symmetry, we construct the model of dense matter using the DDME2 parametrization in the nucleonic sector with the DDRH model and then use that matter to construct the stellar structure of NS. 

\subsection{Matter properties}

Since antikaons are bosons and occupy the $\mathbf{p}=0$ state, their presence suppresses the lepton population in high-density regimes \cite{1994NuPhA.572..693T}.\\
 The threshold densities of $K^-$ and ${\bar{K}^0}$ condensates for different coupling schemes are listed in Table~\ref{table:su6}. 
 As shown in Table~\ref{table:su6}, the threshold density for $K^-$ condensation increases with larger values of $\alpha_m$, which is attributed to the corresponding increase in the chemical potential $\mu_{K^-}$. For $z_b=2z_m$ and $z_b=z_m/2$, the threshold density of $K^-$ does not change with an increase in $z_m$ value, while for $z_b=z_m$, the threshold density for $K^-$ increases with $z_m$. 

\begin{figure*}[!ht]   
    \centering
    \includegraphics[angle=-90, width=18cm]{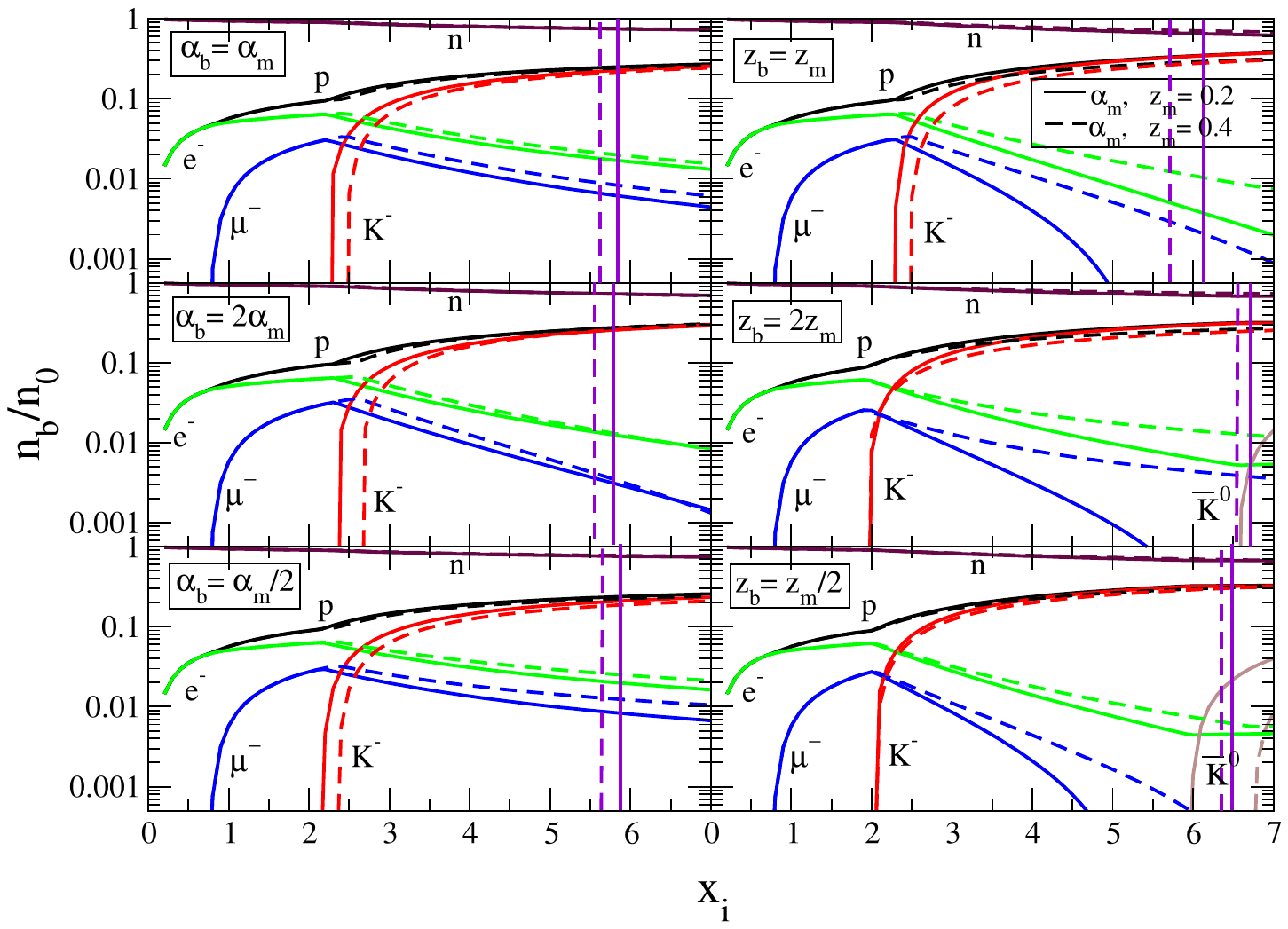}
    \caption{Color online: The variation of particle fractions $x_i$ with the normalized baryon number density in NK matter for $U_{\bar K} = -130$ MeV. Left panels: for different $\alpha_m$. The upper panel corresponds to $\alpha_b = \alpha_m$, the middle panel to $\alpha_b = 2\alpha_m$, and the lower panel to $\alpha_b = \alpha_m/2$. Right panels: for different $z_m$. The upper panel shows the case $z_b = z_m$, the middle panel $z_b = 2z_m$, and the lower panel $z_b = z_m/2$. The solid lines and the dashed lines are for $\alpha_m,z_m = 0.2$ and $0.4$, respectively. The vertical lines denote the central matter densities corresponding to the maximum mass NS configuration for the respective EOS models. }
    \label{fig:abundances}
\end{figure*}

\begin{figure*}[!ht]   
    \centering
    \includegraphics[angle=-90, width=12cm]{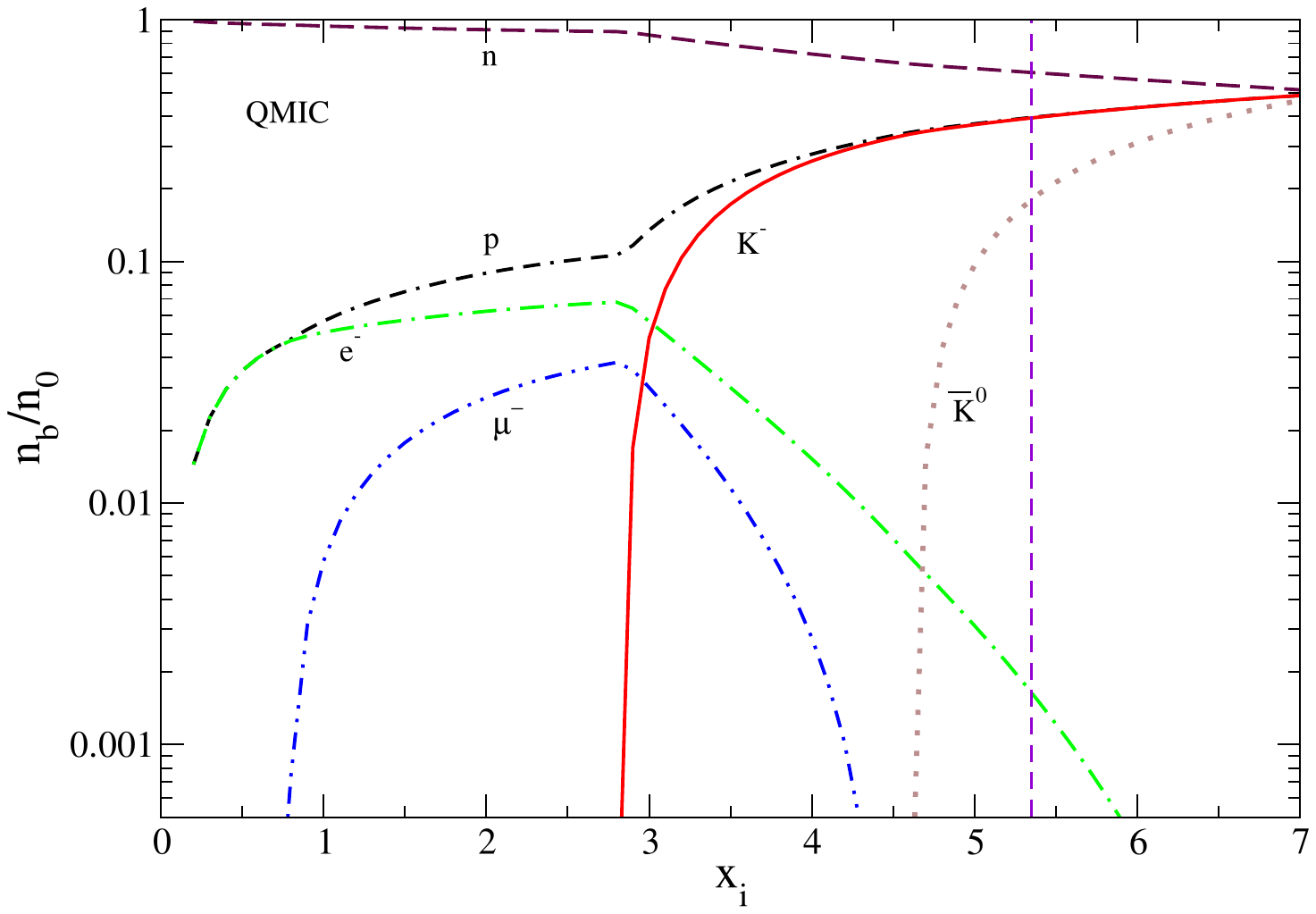}

    \caption{Color online: The variation of particle fractions $x_i$ with normalised baryon number density in NK matter for the case of QMIC case for $U_{\bar K}=-130$ MeV. The vertical line denotes the same fact as in Fig. \ref{fig:abundances}.}
    \label{fig:abundance_QMIC}
\end{figure*}

The relative abundances of different particle species for $\alpha_m = 0.2$ and $0.4$, for the optical potential value of $U_{\bar{K}} = -130$~MeV, are shown in Fig.~\ref{fig:abundances} for the three cases of $\alpha_b= \alpha_m, 2\alpha_m, \alpha_m/2$ in the left panel and the relative abundances for $z_m = 0.2$ and $0.4$, with the same optical potential for the cases of $z_b= z_m, 2z_m, z_m/2$ in the right panel of the figure. It is evident from Fig.~\ref{fig:abundances} and Table~\ref{table:su6} that the threshold density for $K^-$ condensation increases with increasing $\alpha_m$ in the case of varying $\alpha_m$ scenario. For comparison, the particle abundances for QMIC are shown in Fig. \ref{fig:abundance_QMIC}.
On comparing with Fig. \ref{fig:abundances}, it is noteworthy to mention that the onset of $K^-$ condensation is observed at a higher matter density in the case of QMIC.
Furthermore, $\bar{K}^0$ does not appear within the stable dense matter in most of the EOSs considering SU(3) couplings. This distinction can be attributed to the substantially large values of the $g_{\phi \bar{K}}$ coupling.

\begin{figure*}   
    \centering
    \includegraphics[angle=-90, width=18cm]{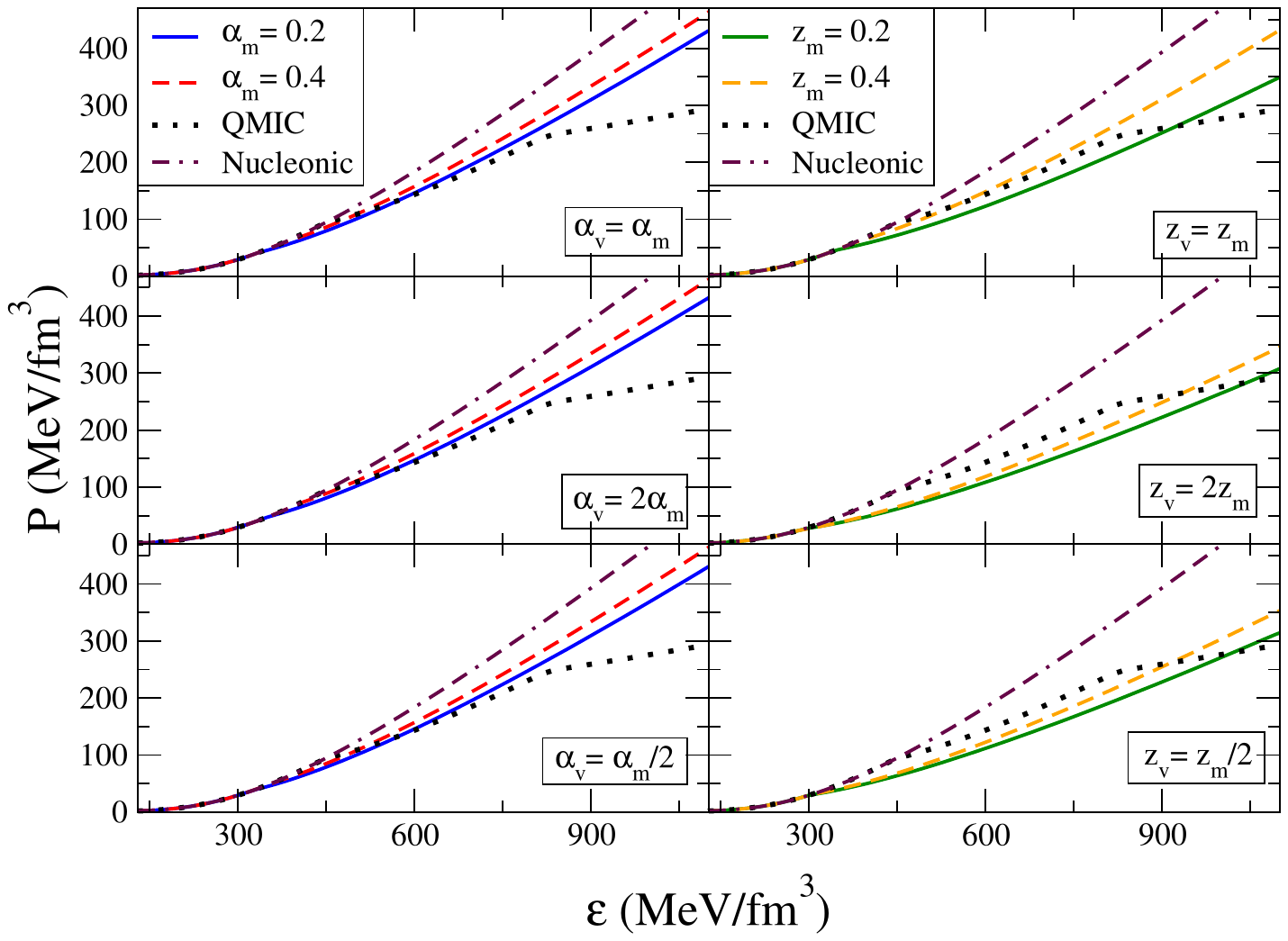}
    \caption{Color online: The variation of pressure with energy density in NK matter for $U_{\bar K} = -130$ MeV. Left panels: for different $\alpha_m$. The upper panel corresponds to $\alpha_b = \alpha_m$, the middle panel to $\alpha_b = 2\alpha_m$, and the lower panel to $\alpha_b = \alpha_m/2$. Right panels: for different $z_m$. The upper panel shows the case $z_b = z_m$, the middle panel $z_b = 2z_m$, and the lower panel $z_b = z_m/2$. The solid line represents $\alpha_m, z_m = 0.2$, the dashed line represents $\alpha_m, z_m = 0.4$, dotted and double-dashed lines represent QMIC and the case of pure nucleonic matter, respectively.}
    \label{fig:EvsP}
\end{figure*}

 We show the matter EOS in Fig. \ref{fig:EvsP} for $\alpha_m = 0.2$ and $0.4$, for the optical potential value of $U_{\bar{K}} = -130$~MeV with the three cases of $\alpha_b= \alpha_m, 2\alpha_m, \alpha_m/2$ in the left panel and the EOS for $z_m = 0.2$ and $0.4$, with the same optical potential for the cases of $z_b= z_m, 2z_m, z_m/2$ in the right panel of the figure. It is observed that the matter EOS becomes stiffer with larger values of $\alpha_m$ when varied at fixed $z_m$, and with larger values of $z_m$ when varied at fixed $\alpha_m$. The EOS that is obtained when we consider only nucleons without antikaon condensation in the matter is shown in the Fig. \ref{fig:EvsP}.
 
Hence, for all the cases, the initial kink represents the appearance of $K^-$ condensate. The antikaon condensation in the matter occurs through a second-order phase transition.
In this context, the order parameter of the transition is the condensate density, $n_{K^-}$, which remains zero below the threshold and increases continuously from zero once the condition $\mu_e=\omega_{K^-}$ is satisfied. This continuous growth of the order parameter, while the pressure and baryon chemical potential remain smooth across the transition, is characteristic of a second-order phase transition \cite{Muto:2005my, PhysRevC.62.035803}.
It is observed that the matter is stiffer with the SU(3) antikaon coupling scheme compared to that with QMIC in all the scenarios except in the cases where we consider the condition of $z_b = 2z_m$ and $z_b = z_m/2$ and at $z_m=0.2$ for $z_b = z_m$.

\begin{table*}  

\caption{Maximum mass, $M_{\text{max}}$ (in units of $M_\odot$), corresponding central density and threshold densities for antikaon condensation for the value of optical potential depth $U_{\bar{K}} = -130$ MeV at $n_0 = 0.152\, \text{fm}^{-3}$ for the case of varying $\alpha_m$ (left panel) and varying $z_m$ (right panel).}
\centering
\resizebox{1.1\textwidth}{!}{%
\begin{tabular}{cccccc|cccccc}
\hline
\multicolumn{6}{c}{$\alpha_m$ varying} & \multicolumn{6}{c}{$z_m$ varying}\\
\hline
 & $\alpha_m$ & $x^{K^{-}}_{th}$ & $x^{K^{0}}_{th}$ & $M_{\text{max}}$ & $x_{\text{central}}$ &
 & $z_m$ & $x^{K^{-}}_{th}$ & $x^{K^{0}}_{th}$ & $M_{\text{max}}$ & $x_{\text{central}}$ \\
\hline
$\alpha_b=\alpha_m$   & 0.2 & 2.30 & -- & 2.26 & 5.84 & $z_b=z_m$ &0.2 & 2.30 & -- & 2.09 & 6.13 \\
& 0.4 & 2.50 & -- & 2.34 & 5.61 &  &0.4 & 2.50 & -- & 2.28 & 5.71 \\
&1.0 & 3.50& -- & 2.42 & 5.31 &  &0.8 & 3.10 & -- & 2.45 & 5.37 \\
\hline
$\alpha_b=2\alpha_m$  & 0.2 & 2.40 & -- & 2.27 & 5.79 & $z_b=2z_m$ &0.2 & 2.00 & 6.60 & 1.94 & 6.72 \\
& 0.4 & 2.70 & -- & 2.35 & 5.55 & &0.4 & 2.00 & --   & 2.04 & 6.56 \\
& 0.5 & 2.80 & -- & 2.38 & 5.47 & &-- & -- & --   & -- & -- \\
\hline
$\alpha_b=\alpha_m/2$ & 0.2 & 2.20 & -- & 2.26 & 5.87 & $z_b=z_m/2$ &0.2 & 2.10 & 5.60 & 1.97 & 6.54 \\
& 0.4 & 2.40 & -- & 2.33 & 5.64 & &0.4 & 2.10 & 6.80 & 2.07 & 6.35 \\
& 1.0 & 3.40 & -- & 2.46 & 5.35 & &0.8 & 2.20 & -- & 2.25 & 5.92 \\
\hline
QMIC                  & --  & 2.90 & 4.70 & 2.24 & 5.35 & QMIC& -- & 2.90 & 4.70 & 2.24 & 5.35 \\
\hline 
Nucleonic Case & --  & -- & -- & 2.48 & 5.34 & Nucleonic Case& --  & -- & -- & 2.48 & 5.34 \\
\hline 
\end{tabular}
}
\label{table:su6}

\end{table*}

\subsubsection{Stellar structure}

\begin{figure*}   
    \centering
    \includegraphics[angle=-90,width=18cm]{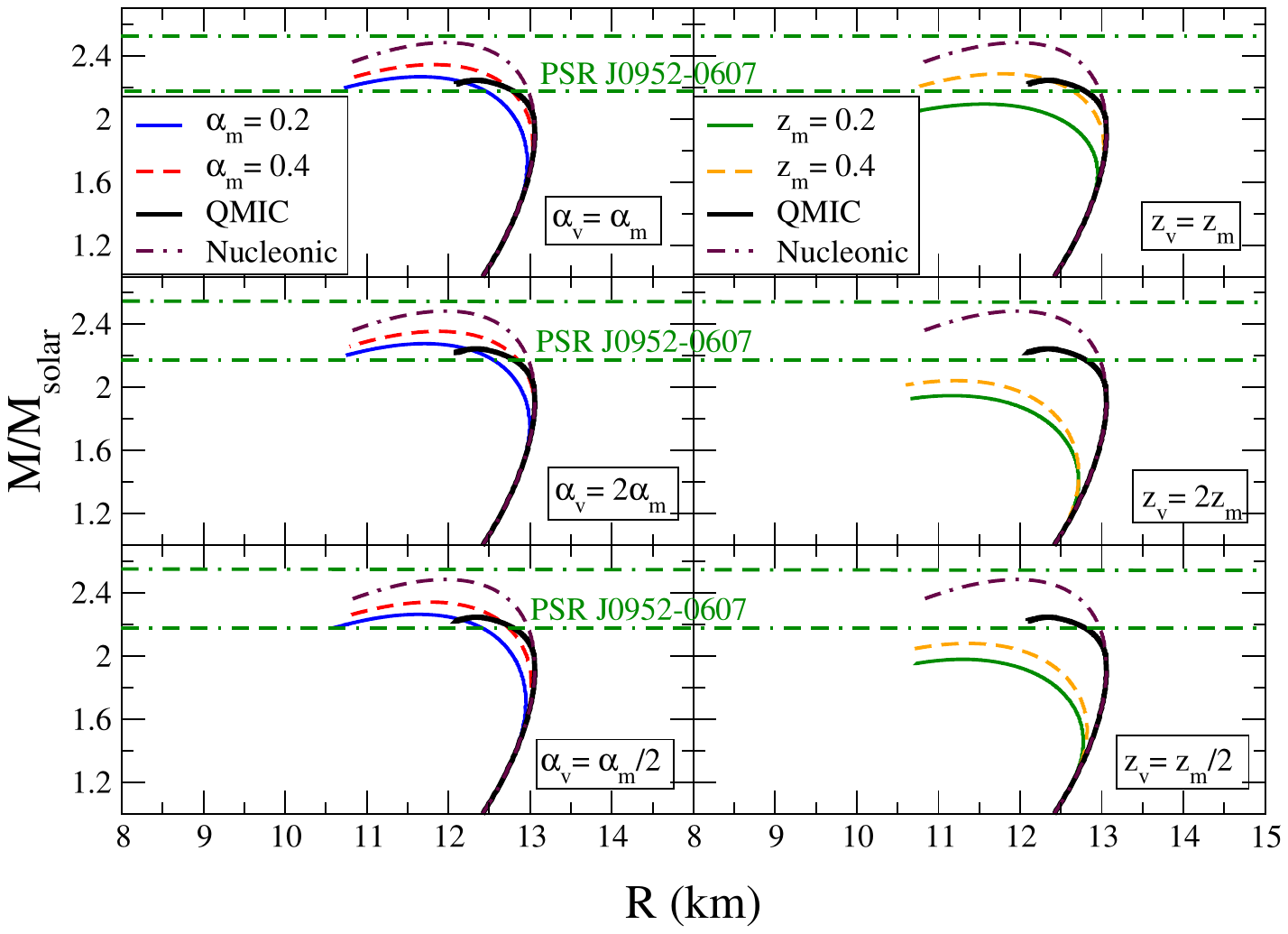}
    \caption{Color online: The variation of mass with radius in NK matter for $U_{\bar K} = -130$ MeV. The observed mass included in the figure is of PSR J0952-607 (M = 2.35 ± 0.17$M_\odot$ ) \cite{2022ApJ...934L..17R}. Left panels: for different $\alpha_m$. The upper panel corresponds to $\alpha_b = \alpha_m$, the middle panel to $\alpha_b = 2\alpha_m$, and the lower panel to $\alpha_b = \alpha_m/2$. Right panels: for different $z_m$. The upper panel shows the case $z_b = z_m$, the middle panel $z_b = 2z_m$, and the lower panel $z_b = z_m/2$. The solid line represents $\alpha_m, z_m = 0.2$, the dashed line represents $\alpha_m, z_m = 0.4$, thick solid black, and the double-dashed lines represent QMIC and the case of pure nucleonic matter, respectively.}
    \label{fig:MvsR}
\end{figure*}

The structure of a spherically symmetric, non-rotating star in hydrostatic equilibrium under the influence of gravity is described by the Tolman–Oppenheimer–Volkoff (TOV) equations \cite{1939PhRv...55..364T, 1939PhRv...55..374O, 1997A&A...328..274B}. The mass–radius (M–R) relation for static stars is obtained by solving these equations, with the Baym–Pethick–Sutherland (BPS) EOS adopted for the crust \cite{1971ApJ...170..299B}. Considering the initial case of fixed $z_m$ and varying $\alpha_m$, is shown in the left panel of Fig.~\ref{fig:MvsR} for the three different cases of $\alpha_b$ at $U_{\bar{K}}=-130$ MeV. As $\alpha_m$ increases, the vector meson contribution to the antikaon coupling becomes more significant, leading to a stiffer EOS. Consequently, the maximum mass of the star also increases. For the case of fixed  $\alpha_m$ and varying $z_m$, the M–R relation is presented in the right panel of Fig.~\ref{fig:MvsR} for the three different cases at $U_{\bar{K}}=-130$ MeV. An increase in the parameter  $z_m$, which modulates the relative strength of the singlet and octet couplings in the meson–antikaon interaction, leads to a stiffer EOS. As a result, the corresponding stellar models attain higher maximum masses. The M–R relation for the case without antikaon condensation is presented in Fig.~\ref {fig:MvsR} for comparison.

Also, as matter EOS in the scenario of varying $\alpha_m$ is stiffer compared to the QMIC scheme, the maximum achievable mass for the SU(3) coupling scheme is higher than that for the QMIC case. In contrast, for the scenario of varying $z_m$, only the cases with $z_m>0.4$ under the condition $z_b=z_m$ are stiffer than QMIC. 

It is noted from Table \ref{table:su6} that the threshold of ${\bar{K}^0}$ lies beyond the central density of the maximum mass star, except for the cases with $z_b = z_m/2$ and at $z_m = 0.2$ for $z_b = 2z_m$. Therefore, ${\bar{K}^0}$ condensation does not occur within the star in all other cases.

The maximum mass and corresponding central energy density of the star for all the cases are tabulated in Table \ref{table:su6} with relevant parameters. The calculated maximum masses are in good agreement with the observational constraint set by the massive pulsar PSR J0952–0607, with a measured mass of $M = 2.35 \pm 0.17~M_\odot$~\cite{2022ApJ...934L..17R} except in the cases where we consider the condition of $z_b = z_m$ at $z_m=0.2$, $z_b = 2z_m$ and $z_b = z_m/2$. This is illustrated in Fig.~\ref{fig:MvsR}. 
The M–R relation for the case of pure nucleonic matter is also shown in Fig.~\ref{fig:MvsR}. In the absence of antikaon condensation, the corresponding matter EOS is noticeably stiffer than in the cases with antikaon-condensed matter, leading to higher maximum masses. This highlights the significant softening effect of antikaon condensation on the EOS.

\section{Summary and conclusion}\label{sec:conclu}
We have calculated couplings of antikaons with vector mesons such as $\omega$, $\rho$, and $\phi$ in flavour SU(3) symmetry — marking the first time this symmetry has been applied to the study of antikaon condensation in dense matter. These couplings depend on the parameter $\alpha_v$, which represents the weight factor for the symmetric ($D$) and antisymmetric ($F$) contributions and varies between $0$ and $1$, as well as on the parameter $z$, defined as the ratio of the octet coupling ($g_8$) to the singlet coupling ($g_1$), which ranges from 0 to $2/\sqrt{6}$.

We considered two scenarios, each comprising three cases. In the first scenario, the value of $z_m$ is fixed while $\alpha_m$ is varied within its allowed range. This includes the cases $\alpha_b = \alpha_m$, $2\alpha_m$, and $\alpha_m/2$, all evaluated under the condition $z_b = z_m$. In the second scenario, $\alpha_m$ is fixed while $z_m$ is varied throughout its range, for the three cases $z_b = z_m$, $2z_m$, and $z_m/2$, with the condition $\alpha_b = \alpha_m$. The scalar $\sigma$-antikaon coupling is determined using the antikaon potential depth $U_{\bar{K}} = -130$ MeV in normal nuclear matter at saturation density $n_0 = 0.152\,\mathrm{fm}^{-3}$ for all the cases. We have utilised these antikaon couplings with scalar and vector mesons to investigate the compositions, EOS, and mass-radius relations in baryonic matter transitioning to antikaon-condensed matter through a second-order phase transition. For this analysis, we employed the DDME2 parametrization for nucleon-meson couplings.

Building on this analysis, we examined the EOSs for dense matter using the previously mentioned value of $U_{\bar K}$. These results are illustrated in Fig. \ref{fig:EvsP}.
 As the values of $\alpha_m$ and $z_m$ increase in the two separate scenarios, the EOS becomes progressively stiffer. The matter EOS obtained using SU(3) couplings is stiffer compared to the QMIC scheme for the cases of varying $\alpha_m$, and for the cases with $z_m > 0.4$ under the condition $z_b = z_m$ in the varying $z_m$ scenario, resulting in a higher maximum attainable mass for compact stars in the SU(3) case. A notable outcome of our analysis is that the configurations with $\alpha_m = 1$ (for fixed $z_m$) and $z_m = 0.8$ (for fixed $\alpha_m$), together with the cases where the baryonic parameter is taken to be twice the mesonic parameter, i.e., $\alpha_m = 0.5$ (for fixed $z_m$) and $z_m = 0.4$ (for fixed $\alpha_m$), represent the upper limits of the parameters in their respective cases and yield the stiffest EOS among all the configurations considered.

\section*{Data Availability}
The data used in the manuscript can be obtained at reasonable request from the corresponding author.

\section*{Acknowledgements}
The authors thank the anonymous referee for the valuable suggestions provided, which have greatly contributed to enhancing the quality and presentation of this manuscript. The authors acknowledge the financial support from the Science and Engineering Research Board (SERB), Department of Science and Technology, Government of India, through project no. CRG/2022/000069. MS acknowledges the partial financial support from the DRDO through project no. DGTM/ERIP/GIA/24-25/010/005. 
The authors would like to thank Debades Bandyopadhyay for his enlightening discussions and recommendations.

\bibliography{references}
\bibliographystyle{unsrt}
\end{document}